\documentstyle[12pt,epsf]{article}
\newcommand{\be}{\begin{eqnarray}}
\newcommand{\ee}{\end{eqnarray}}

\catcode`\@=11
\def\lsim{\mathrel{\mathpalette\@versim<}}
\def\gsim{\mathrel{\mathpalette\@versim>}}
\def\@versim#1#2{\vcenter{\offinterlineskip
\ialign{$\m@th#1\hfil##\hfil$\crcr#2\crcr\sim\crcr } }}
\catcode`\@=12

\begin{document}

\begin{center}
{\Large\bf Running couplings in extra dimensions}
\end{center} 
\vspace{0.5cm}
\begin{center}
{\sc Jisuke Kubo} and {\sc Haruhiko Terao}
\footnote{Parallel session talk presented at ICHEP 2000, 
July 27 - August 2, 2000, Osaka, Japan
}
\\
{\em 
Institute for Theoretical Physics, 
Kanazawa  University, 
Kanazawa 920-1192, Japan}\\
E-mail: jik@hep.s.kanazawa-u.ac.jp, 
terao@hep.s.kanazawa-u.ac.jp
\vspace{2mm}\\
{\sc George Zoupanos}\\
{\em
Physics Department, Nat. Technical University, 
GR-157 80, Zografou, Athens, Greece}\\
E-mail: George.Zoupanos@cern.ch 
\end{center}

\vspace{5mm}
\begin{center}
{\sc\large Abstract}
\end{center}

\noindent
The regularization scheme dependence of
running couplings in extra compactified dimensions is
discussed. We examine several regularization schemes 
explicitly in order to analyze the
scheme dependence of the Kaluza-Klein threshold effects,
which cause the power law running,
in the case of the scalar theory in five dimensions with
one dimension compactified.
It is found that in 1-loop order, 
the net difference in the running of the coupling
among the different schemes is reduced to be rather small
after finite renormalization.
An additional comment concerns the running couplings in the
warped extra dimensions which are found to be regularization
dependent above TeV scale.

\section{RG in large extra dimensions}
Recently the extra compactified dimensions have been attracting
much attention as possibilities to explain various hierarchy
problems; the gauge hierarchy, the Yukawa hierarchy and so on.
The effective field theories in extra dimensions contain 
towers of massive Kaluza-Klein (KK) excitations, whose quantum 
effects alter the behavior of the running couplings from 
logarithmic to power. Therefore the traditional picture of gauge
coupling unification may be drastically changed \cite{ddg}.
In this talk we examine the running coulings explicitly
in several schemes and consider the origin of power law
running and their scheme (in)dependence \cite{ktz}. 

In practice the notion of running couplings is not 
well defined in extra dimensions due to nonrenormalizability. 
We cannot help but defining the field theories by a 
certain cutoff. Then the Wilson RG is supposed to offer 
a natural framework to define the RG flows in such cases. 
There is another reason why the Wilson RG is suitable for
the calculation of the $\beta$-functions in extra dimensions.
The power law behavior of the running coupling is supposed
to be generated by successive threshold corrections by tower
of massive KK modes. The Wilson RG is faithful to
their decoupling effects.

Therefore we apply the Exact RG, which is a 
continuum formulation of the Wilson RG, to a scalar field theory in
$M_4 \times S^1$ space-time. In this formulation a certain cutoff
is performed to the internal loop momenta, and the running
coupling is defined by variation of the cutoff scale.
We derive the $\beta$-functions for the four scalar coupling in the 
1-loop level, assuming that the coupling is weak enough.
We also calculate the $\beta$-functions in other schemes: the proper
time regularization and the momentum subtraction, and compare
these results. 

In all schemes we obtain the $\beta$-functions as
\begin{equation}
\beta_{\lambda} = \frac{d\lambda}{dt} = b \epsilon_k(t)\lambda^2,
\end{equation}
where scale parameter $t=\ln R\Lambda$ is introduced in terms of
radius of the compact space $R$ and the cutoff scale $\Lambda$.
$b$ is the 1-loop coefficient in four dimensions.  
The function $\epsilon_k(t)$ is dependent on each scheme $k$, and 
it's asymptotic form is given by
\begin{equation}
\begin{array}{lclll}
\epsilon_k(t) & \rightarrow & 1 & {\rm for} & t \ll 0, \\ 
              & \rightarrow & B(k)e^t & {\rm for} & t \gg 0, 
\end{array}
\end{equation}
where $B_k$ is a scheme dependent constant.
This function is shown in Fig.1 by truncating the KK
modes at $N$-th level. 
\begin{figure}
\epsfxsize=0.5\textwidth
\centerline{\epsffile{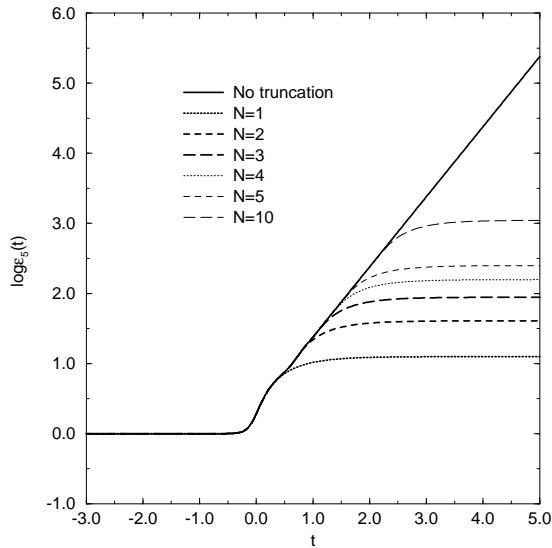}}
\caption{Scale dependence of the $\beta$-function coefficient calculated in
an ERG scheme.}
\label{fig1}
\end{figure}
It is seen that the $\beta$-function
shifts to the five dimensional one smoothly
by the KK threshold corrections. 
We may define the effective dimensions \cite{ktz} by
\begin{equation}
D_{\rm  eff}(t) = 4 + \frac{d \ln \epsilon_k}{dt},
\end{equation}
which shows smooth transition from 4 to 5.

\section{Scheme dependence}

The $\beta$-functions given in Eq.~(1) are scheme dependent
even in the 1-loop level. However we may redefine the
coupling constants in different schemes so that the 
$\beta$-functions match in the asymptotic region. 
The scheme dependence remained even after this procedure
represents the net ambiguity in the RG in extra dimensions.
The results obtained by explicit calculations are shown in
Fig.~2. It is seen that the unremovable scheme dependence
is rather small. In consequence, it is found that the 
GUT predictions for the low energy gauge couplings
\cite{ddg} are almost scheme independent \cite{ktz}.

\begin{figure}
\epsfxsize=0.5\textwidth
\centerline{\epsffile{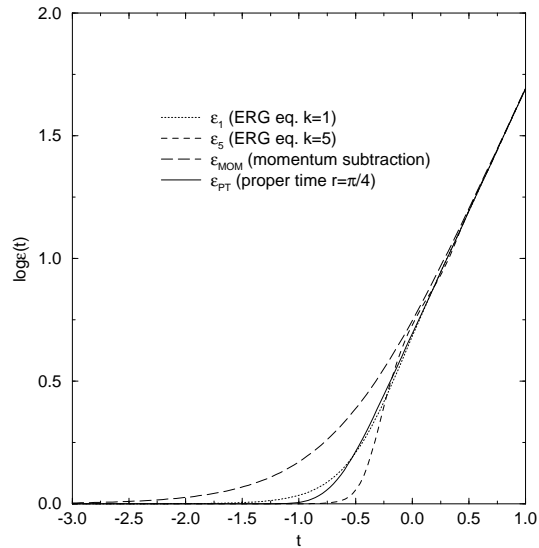}}
\caption{Scheme dependence of the net thershold corrections.}
\label{fig3}
\end{figure}

\section{RG in the warped extra dimensions}

The Randall-Sundrum type of compactification \cite{rs} 
is now under the most vigorous investigation. 
The extra dimension is bounded by two three
branes, in one of which the Planck scale is reduced to TeV scale.
A peculiarity of this compactification is 
that the mass scale of the KK tower appears at TeV order, even if
the bare mass is Planck scale \cite{gw}.
If we apply the RG scheme mentioned above to this case, the
running couplings are found to show power law behavior too.
This should be compared with the results obtained by the PV 
regularization \cite{pomarol}.
However it should be said that the running coupling above TeV 
scale is totally regularization dependent, since the KK 
spectrum is influenced by the Planck scale (string) physics.

\end{document}